\documentclass[aps,pra,twocolumn,groupedaddress,showpacs,amsmath,amsfonts,floatfix]{revtex4}
\usepackage{graphicx}
\usepackage{bm}
\usepackage{color}
\usepackage{relsize}

\begin{document}
\title{The Effect of Strained Bonds on the Electronic Structure of Amorphous Silicon }

\author{Reza Vatan Meidanshahi}
\email{rvatanme@asu.edu}
\author{Payam Mehr}
\author{Stephen M. Goodnick}
\affiliation{School of Electrical, Computer and Energy Engineering, Tempe, AZ, 85281, USA}

\date{\today}

\begin{abstract} 
Several amorphous silicon structures were generated using a classical molecular dynamics (MD) protocol of melting and quenching with different quenching rates. An analysis of the calculated electronic properties of these structures revealed that the midgap state density of a-Si which is of interest for solar cell and thin film transistor applications can be correlated to bond angle standard deviation. We also found that this parameter can strongly determine the excess energy of a-Si, which is an important criteria in theoretically generating realistic atomic structures of a-Si.
\end{abstract}

\pacs{Valid PACS appear here}

\maketitle
\section{\label{sec:intro}Introduction}

Amorphous silicon (a-Si) has been extensively investigated as the archetypal amorphous covalently bonded material and is widely used in numerous electronic and photovoltaic devices. These features have spurred an intense interest in their electronic properties over the last few decades. a-Si can be fabricated using different methods like laser melting, ion implantation, and growth techniques \cite{laaziri1999high, kail2011configurational, fuentes2017fabrication}. The atomic structure of the a-Si obtained using different fabrication methods to a large extent depends on the preparation technique. As an example, ion implantation often gives samples with large dangling bonds (DBs) and vapor deposition results in samples with voids. These differences in atomic structure as a result of the fabrication method, strongly affects the electronic properties of a-Si. Though it has been extensively investigated, the exact nature of the relation, between atomic structure and electronic properties, has been a subject of intense contention. In addition to gaining a more accurate understanding of the basic characteristics of covalently bonded amorphous materials, a better understanding of this relation would be helpful for technological progress.

	One of the main electronic features which makes a-Si different from its crystalline counterpart is the presence of electronic states in its energy gap which are called midgap states. The density of midgap states can control many electronic properties of the material by affecting trapping and recombination processes and consequently affect device functionality \cite{mehr2018soi}. The common belief is that midgap states are generated by coordination defects or DBs. However, there is ample evidence that other less unstable structural defects might also contribute to these states \cite{shimizu2004staebler, gotoh1998experimental, stratakis2002photoinduced, wagner2008microscopic, khomyakov2011large, pfanner2013dangling, meidanshahi2019electronic, meidanshahi2020defect}. Calculations of defects in a-Si with or without hydrogen, were mostly involved in understanding its physical behavior. Nevertheless, significant recent developments in the theoretical analyses of the defect-midgap state relation took place were made by investigating the electronic structure of several configurations of a-Si or a-Si:H. Wagner et al. \cite{wagner2008microscopic} created different amorphous structures by applying Wooten-Wiener-Weaire process with the Keating potential and studied their electronic properties within a density-functional theory (DFT) approach. They observed that configurations with high strained bonds (SBs), as the configuration with DBs, contain highly dense midgaps which are able to trap holes in a-Si. Khomyakov et al. \cite{khomyakov2011large} created a-Si:H model of 500 atoms by applying large-scale replica-exchange MD using DFT-derived classical potentials. Their bond-resolved density of states (DOS) indicated that, contrary to common belief, domains with highly strained Si-Si bonds significantly contribute to midgap states density no less than DBs. Using a first principle study of electron paramagnetic resonance (EPR), Pfanner et al. \cite{pfanner2013dangling} showed that a strong indirect effect of network and strained bonds on creating midgap states. In addition to a-Si, the relative importance of SBs and DBs in determining the midgap state density is relevant for other amorphous materials also. However, except for the correlation between midgap state and DBs, none of the previous investigations have provided a quantitative relation between midgap state density and other structural parameters.

	This paper presents quantitative insights on the relation between the midgap state density and several structural parameters. The specific aim of this paper is twofold: (i) to identify which of the structural defects including dangling bonds (DBs), floating bonds (FBs), bond length average (BLA), bond length standard deviation (BLSTD), bond angle average (BAA) and bond angle standard deviation (BASTD) are capable of accurately describing the nature of midgap state density (ii) to clarify the role of these structural defect on the excess energy of a-Si, which has been recently proposed as an important electronic property in the simulation of a-Si \cite{pedersen2017optimal}. In our approach, the first step involves the generation of 23 different large supercells of a-Si models with 216 Si atoms. In this step, molecular dynamics simulation of melting and quenching process with different cooling rate was applied to crystalline silicon (c-Si) supercell while the inter-atomic interaction is described by the Tersoff potential. In the second step, the structures are optimized by carrying out DFT relaxation calculations. In the third step, the integrated density of midgap states, the excess energy of obtained structures are computed on the relaxed structures using DFT simulations. These quantities are then correlated to each of the aforementioned structural parameters. Finally, the calculated results are compared to previous computational investigations. 

\section{Method}
\subsection{Technical Details}

We use a melting and quenching approach to generate a structural model of a-Si as the starting atomic structure for the DFT calculations. The LAMMPS molecular dynamic code \cite{plimpton1995fast} is used for simulating the melting and quenching processes. In the MD simulations, the Tersoff interatomic potential \cite{tersoff1989modeling} was employed for describing Si atom interactions, with a cut-off radii of 2.7 \AA \space (taper) and 3.0 \AA \space (maximum); this potential has been widely used for generating Si based structures \cite{ohira1994molecular, nolan2012surface, legesse2014first}. Full ion relaxation of the resulting structure from the MD simulation was performed at the DFT-level as implemented in the Quantum Espresso 6.2.1 software package \cite{giannozzi2009quantum}. The BFGS quasi-Newton algorithm, based on the trust radius procedure was used as the optimization algorithm to find the relaxed structure. The structural analysis of the final a-Si structure was performed using the ISAACS software \cite{le2010isaacs}. 
	Both ionic relaxation and electronic structure calculations were performed using the Becke-Lee Yang-Parr (BLYP) exchange-correlation functional \cite{becke1993density, lee1988development}. The core and valence electron interactions were described by the Norm-Conserving Pseudopotential function. Unless otherwise stated, an energy cutoff of 12 Ry was selected for the plane-wave basis set.  A 4$\times$4$\times$4 k-point mesh with the Monkhorst-Pack grids method for the Brillouin-zone sampling was employed in all the calculations. To determine the band occupations and electronic density of states fixed method was exploited.

\subsection{Generation of a-Si:H/c-Si Structures}

Molecular dynamics simulations in conjunction with DFT calculations have been demonstrated to yield amorphous material structures whose properties are commensurate with experimental results \cite{ishimaru1997generation, vstich1991amorphous}. Therefore, we initially carried out MD simulations to generate a general form of the a-Si structure, and then relaxed the structure using a DFT calculation to obtain experimentally compatible structures. MD simulation of the melting and quenching process was carried out on a crystalline Si structure in order to create an a-Si supercell containing 216 Si atoms (a-Si216) with three dimensional (3D) periodic boundary conditions. A diamond starting atomic structure of crystalline Si with a lattice constant of a\textsubscript{0} = 5.46 \AA \space was constructed using a cubic supercell with the dimension of a = b = c = 3a\textsubscript{0} , which was periodically repeated in 3-D space to generate an infinite network of atoms. The value of a\textsubscript{0} was chosen in such a way that the mass density of our supercell is equal to the mass density of a-Si measured by experiments \cite{custer1994density, smets2003vacancies}.  Then, we simulated a 10 ps melting process at 3000 K in 0.1 fs time steps, with a fixed volume and temperature ensemble (NVT).The structure was then quenched to 300 K with different cooling rates ranging from 9$\times$10\textsuperscript{10} to 3$\times$10\textsuperscript{14} K/s,  and annealed for 25 ps at 300 K afterwards. Finally, the structure was optimized using a DFT relaxation calculation.

\begin{figure}[tb]
\includegraphics[width=60mm]{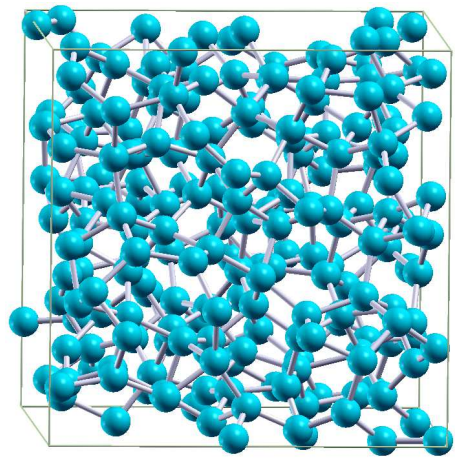}
\caption{ The atomic structure of a typical simulated a-Si216 supercell.
}\end{figure}

Figure 1 illustrates the atomic structures of a typical simulated a-Si supercell containing 216 Si atoms obtained from MD simulation with a cooling rate of ~10\textsuperscript{11} and DFT relaxtion calculation. Based on the periodic structure formed from the supercell in Figure 1, we found one dangling bond and one floating bond per supercell, with an assumed Si-Si bond length cutoff of 2.58 \AA, which is 10\% longer than the experimental Si-Si bond length (2.35 \AA). The structure mostly displays stable 5, 6, or 7 fold rings, and there are no large voids or holes in it. The average Si-Si bond length is 2.354 \AA \space with an rms value of 0.049 \AA. The average Si-Si-Si bond angle is 108.2$^{\circ}$ with an rms value of 13.7$^{\circ}$. In the crystalline form of Si, the Si-Si bond length is 2.35 \AA \space and the Si-Si-Si bond angle is 109.5$^{\circ}$.

\begin{figure*}[tb]
\includegraphics[width=120mm]{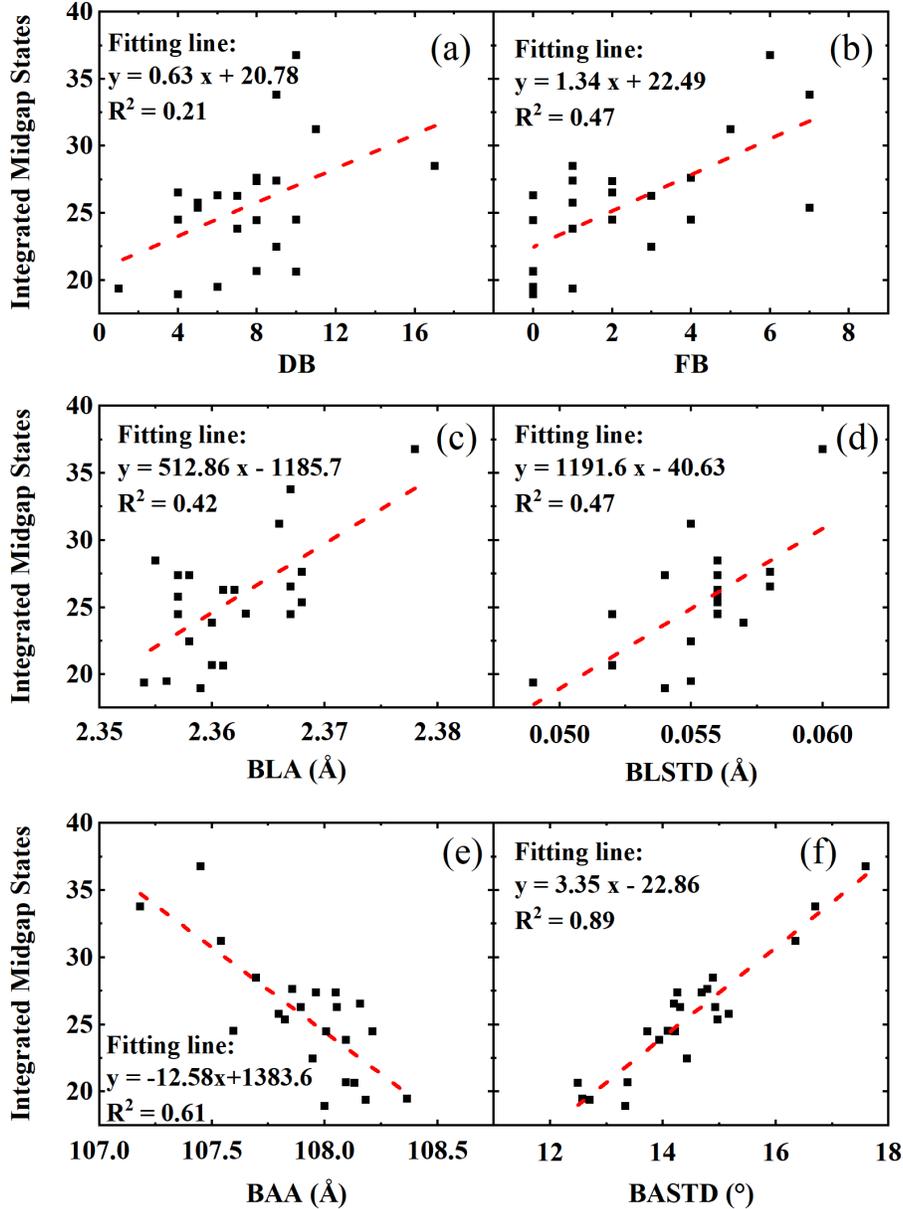}
\caption{ Integrated midgap states vs different structural defects a) dangling bonds (DB) b) Floating Bonds (FB) c) bond length average (BLA) d) bond length standard deviation (BLSTD) e) bond angle average (BAA) and f) bond angle standard deviation (BASTD).
}\end{figure*}

\section{Results}
\subsection{Midgap States Density}
The crystalline form of bulk Si shows clear valence band to conduction band energy gap without any midgap states inside the gap. However, midgap states only exist in the amorphous form of Si. The number of midgap states is highly dependent on the atomic structure of the amorphous network. In order to identify the structural parameter having the most significant effect on the midgap states, the total number of midgap states in the obtained a-Si supercells were calculated by integrating the density of states of a-Si inside the energy range of band-gap of Si in its perfect crystal form. By performing DFT calculations on perfect crystal form of Si, we found that the  energy range of Si band gap in crystalline form is 6.6-7.5 eV.

Fig. 2 illustrates the integrated density of states variations versus different structural parameters. A  linear fit was applied to all the graphs and the obtained regression coefficients were considered as the criteria for determining the accuracy of the fitted and the obtained relations. As Figure 1 denotes, it is obvious that the greatest regression is obtained for the relation of integrated midgap states and bond angle standard deviation. The obtained relation is as follow.

\begin{equation}
I=3.35{\mathlarger{\mathlarger{\sigma}}}_{BASTD}-22.86
\label{IPR-Eq1}
\end{equation}

Where I and ${\mathlarger{\mathlarger{\sigma}}}_{BASTD}$ stand for integrated midgap states and bond angle standard deviation, respectively. According this relation, the integrated midgap states monotonically increases as the bond angle standard deviation increases.
Since any bond angle deviation from the ideal value (109.45$^{\circ}$) cause bond strain, the bond angle standard deviation is an estimation of stored strain in an amorphous structure. Therefore, the obtained dependency of midgap state density and bond angle standard deviation indicates that strained bonds can significantly cause midgap states even more than dangling bond. This finding is in contrast with the common belief that the midgap states are only due to dangling bonds, and is  consistent with the recent studies in the importance of strained bonds in creating midgap states \cite{pfanner2013dangling, meidanshahi2019electronic}.

The low regression values for bond length average and bond length standard deviation denote that none of these quantities are good descriptors for midgap state density of an amorphous structure. As seen from the graphs, bond length average and bond length standard deviation changes is negligble from one structure to another is negligible and therefore the low sensitivity of midgap states density to these quantities is not unexpected. In addition, the obtained regression indicate bond angle average is not a proper quantity for describing the midgap state density of amorphous Si. The reason for this observation might be due to missing the information about the negative and positive bond angle deviation by canceling them each other when they are added up.

Our observation regarding the strong dependency of midgap states density to bond angle standard deviation is reasonable from a chemical bonding perspective. Bond angle standard deviation contains all the information about any deviation, regardless its sign, from ideal bond angle. An ideal bond angle of 109.45 corresponds to the perfect SP\textsuperscript{3} hybridization which cause the bonding and anti-bonding orbitals locate only in the valence band and the conduction band sides, respectively, and no atomic orbital in the band gap. However, when a bond angle associated with a specific atom deviates from its ideal value, the hybridization of that Si atom transform from SP\textsuperscript{3} to SP\textsuperscript{n} where n is an integer number. When a bond angle is greater than 109.45 $^{\circ}$, then n would be less than 3 and consequently some midgap states appear close to the conduction band edge, due to their S-like orbital properties. In vice versa, when a bond angle is less than 109.45 $^{\circ}$, then n would be greater than 3 and consequently some midgap states appear close to the valence band edge, due to their S-like orbital properties.  

The accuracy of the integrated midgap states dependency on the bond angle standard deviation was checked by calculating the integrated midgap states and bond angle standard deviation of the low strained supercell simulated by Pedersen et. al. \cite{pedersen2017optimal}. The mentioned supercell was taken from the reference \cite{pedersen2017optimal} and then was optimized using BLYP functional and finally its electronic density of states was computed. We found that the integrated midgap states and bond angle standard deviation are 11.49 and 10.920 respectively. If we calculate integrated midgap states through the equation 1 using the obtained bond angle standard deviation value, an integrated midgap states of 13.72 will be resulted which is close to the integrated midgap states obtained from DFT calculations (14.15).

\begin{figure*}[tb]
\includegraphics[width=120mm]{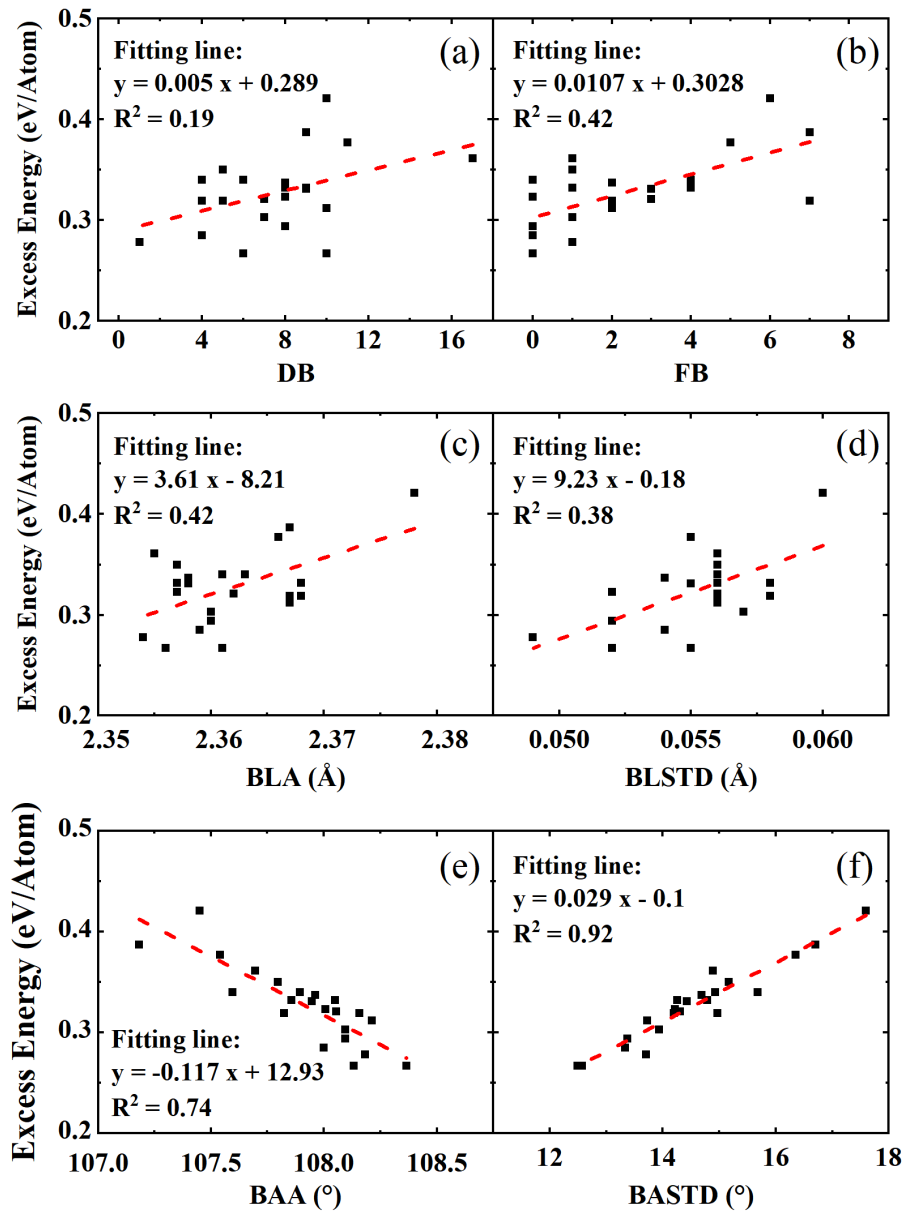}
\caption{ Excess energy vs different structural defects a) dangling bonds (DB) b) Floating Bonds (FB) c) bond length average (BLA) d) bond length standard deviation (BLSTD) e) bond angle average (BAA) and f) bond angle standard deviation (BASTD).
}\end{figure*}

\subsection{Excess Energy}

To further progress in generating an optimal a-Si model, Pederson et al. have recently proposed to focus on the excess energy of the amorphous structure relative to the crystal, a quantity which can be measured from calorimetry experiments \cite{pedersen2017optimal}. This parameter has been getting considered as a critical property in simulation of amorphous structure. Therefore, we also researched the effect of different structural defects on the excess energy stored in an amorphous network.

In order to calculate the excess energy, we initially carried out a DFT relaxation calculation on a crystalline silicon (c-Si) supercell comprised of 216 Si atoms with the same size as the a-Si supercells. Then, we subtracted the energy of a-Si supercell from the energy of c-Si and finally the energy difference is considered as the excess energy stored in the a-Si model. Figure 3 shows the computed excess energies versus different structural defects for the simulated a-Si supercells. As in the midgap states density calculations, a linear regression method is used to fit the resulting points to a line and the regression coefficient is considered as the criteria for the accuracy of the linear relation. It can be concluded from Figure 3 that the best correlation is observed between excess energy and bond angle standard deviation with a regression coefficient of 0.92. As seen from the figure, the obtained equation for calculating excess energy using bond angle standard deviation is as follow.

\begin{equation}
E_{ex}=0.029{\mathlarger{\mathlarger{\sigma}}}_{BASTD}-0.1
\label{IPR-Eq}
\end{equation}

Where E\textsubscript{ex} and ${\mathlarger{\mathlarger{\sigma}}}$ stand for excess energy and bond angle standard deviation, respectively. From equation, it is obvious that the excess energy linearly increases with bond angle standard deviation. The same ration as part a can be used for explaining the strong correlation between E\textsubscript{ex} and ${\mathlarger{\mathlarger{\sigma}}}$. As mentioned before, Si atoms with bond angle deviation from ideal value gets hybridized with SP\textsuperscript{n} instead of SP\textsuperscript{3}. The SP\textsuperscript{n} hybridized orbitals can not create strong equal number of chemical bonds due to their weak overlap compared to SP\textsuperscript{3} orbitals. Hence, regardless of the number of the dangling bonds which is usually considered as the criteria for estimating the stability of a-Si model, the bond angle standard deviation is more important in determining the stability of amorphous network. This finding also shows that the bond angle standard deviation is the more fundamental property of an amorphous network than excess energy that can describe the quality of generated a-Si models. 
Same as part A, we checked the accuracy of our obtained relation using the optimal a-Si model generated by Pedersen et. al. \cite{pedersen2017optimal}. The optimal a-Si structure was taken from the reference and the excess energy was computed after optimizing the structure by BLYP function. Calculating the excess energy by plugging bond angle standard deviation value to equation 2 results in 0.22 eV/atom, which presents a 13.6\% error with respect to the 0.19 eV/atom resulted from the DFT simulations.
\section{CONCLUSION}
Molecular dynamic simulations and DFT relaxation calculations of various a-Si supercells comprised of 216 Si atoms revealed the strong dependency of both integrated density of midgap states and excess energy on the bond angle standard deviation. Consequently, the bond angle standard deviation is more deterministic in the a-Si stability evaluation than the conventional methods in which the number of the dangling bonds are being considered. 

\section{ACKNOWLEDGMENTS}
This material is based upon work primarily supported by the Engineering Research Center Program of the National Science Foundation and the Office of Energy Efficiency and Renewable Energy of the Department of Energy under NSF Cooperative Agreement No.EEC-1041895. Any opinions, findings and conclusions or recommendations expressed in this material are those of the author(s) and do not necessarily reflect those of the National Science Foundation or Department of Energy.

\bibliography{asi}

\end{document}